\documentclass[conference]{IEEEtran}
\usepackage[cmex10]{amsmath}
\usepackage{array}

\usepackage[tight,footnotesize]{subfigure}

\usepackage{times,epsfig}
\usepackage[T1]{fontenc}
\usepackage{graphicx} 
\usepackage{subfigure}
\usepackage{float}
\usepackage{boxedminipage}

\usepackage{cite}

\begin{document}

\title{Joint Estimation and Contention-Resolution \\ Protocol for Wireless Random Access}

\author{\IEEEauthorblockN{\v Cedomir Stefanovi\' c, Kasper F. Trilingsgaard, Nuno K. Pratas and Petar Popovski}
\IEEEauthorblockA{Department of Electronic Systems, Aalborg University, Aalborg, Denmark\\
Email: \{cs,kft88,nup,petarp\}@es.aau.dk}
%\and
%\IEEEauthorblockN{Dejan Vukobratovi\' c}
%\IEEEauthorblockA{
%Department of Power, Electronics and \\ Communication Engineering\\
%University of Novi Sad, Novi Sad, Serbia\\
%Email: dejanv@uns.ac.rs}}
}
\maketitle

\begin{abstract}

We propose a contention-based random-access protocol, designed for wireless networks where the number of users is not a priori known.
The protocol operates in rounds divided into equal-duration slots, performing at the same time estimation of the number of users and resolution of their transmissions.
The users independently access the wireless link on a slot basis with a predefined probability, resulting in a distribution of user transmissions over slots, based on which the estimation and contention resolution are performed.
Specifically, the contention resolution is performed using successive interference cancellation which, coupled with the use of the optimized access probabilities, enables throughputs that are substantially higher than the traditional slotted ALOHA-like protocols.
The key feature of the proposed protocol is that the round durations are not a priori set and they are terminated when the estimation/contention-resolution performance reach the satisfactory levels.

\end{abstract}

\section{Introduction}

Since its inception, slotted ALOHA (SA) \cite{R1975} (and its many variations) has been a popular distributed random access scheme, adopted in many wireless networks.
The common feature for all SA variants is that, in order to maximize the throughput, the knowledge of the expected number of user transmissions per slot is required.
In highly dynamical systems, as is the typical case for wireless access, this parameter is usually a priori not known and it varies over time, hence its estimate has to be obtained for optimal operation.
To this end, a variety of approaches have been proposed, like \cite{Sch1983,V2002,RALRCP2005,KN2006,EL2010}, to name a few.

Building up on the results reported in \cite{SPV2012}, we propose a SA-based random access protocol suited for wireless networks where the number of users is not a priori known.
The protocol operates in rounds that are divided into equal-duration slots.
In every round users access the wireless link on a slot basis with a predefined probability, resulting in a distribution of user transmissions over slots that is used both for estimation and contention-resolution.
Round lengths, measured in number of slots, are not a priori set - they are run until estimation (in the first round) or
contention-resolution (in the following rounds) performance reach satisfactory levels.
Using the combination of the estimation and contention-resolution approaches outlined in the text, we demonstrate that both precise estimates and
high throughputs can be achieved.
A suitable application scenario of the proposed scheme is for wireless access networks with large, but unknown number of devices, as typically encountered in Machine-to-Machine (M2M) scenarios.
The complexity burden of the scheme is on the receiving side, i.e., a Base Station (BS) in M2M scenario, further making it suitable for M2M communications.

The organization of the rest of the paper is as follows.
Section~\ref{sec:background} introduces the background and related work.
The system model and the principles of the proposed protocol are elaborated in Section~\ref{sec:model}.
Section~\ref{sec:results} presents the simulation results, while Section~\ref{sec:conclusion} concludes the paper.

\section{Background and related work}
\label{sec:background}

Framed SA in its basic variant is a simple and straightforward scheme.
Link-time is organized into equal-duration frames, which are divided into equal-duration slots.
The total number of transmissions per frame divided by the number of slots, denoted as the load $G$, is assumed to be constant.
Prior to start of a frame, every user randomly and independently chooses a single slot within that frame in which its transmission is going to take place. Slots can contain no user transmission (i.e., idle slots), just one transmission (singleton slots), or several interfering transmissions (collision slots), and only singleton slots are considered usable.
The well known result for the throughput of framed SA is $T=G e^{-G}$, which attains the maximum $T=1/e \approx 0.37$ when $G = 1$.
To achieve $G=1$ and thus maximize the throughput, the number of slots within the frame should be equal to the total number of transmissions, introducing the need for the estimation of the number of user transmissions.

\begin{figure}[tbp]
	\begin{center}
\includegraphics[width=0.45\columnwidth]{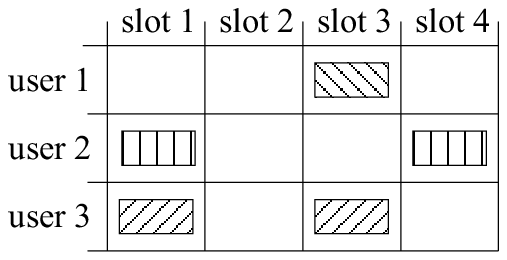}
	\end{center}
\caption{Successive interference cancellation in slotted ALOHA. The transmission of user 2 is resolved in singleton slot 4, enabling the removal of its replica from slot 1 and subsequent resolution of the transmission of user 3 in slot 1. In the same way, resolution of the transmission of user 3 enables the removal of its replica from slot 3, thus resolving the transmission of user 1.}
	\label{fig:SIC}
\end{figure}

Far-reaching changes to the original framed SA were introduced in \cite{CGH2007}, where users were allowed to perform multiple transmissions (i.e., repetitions) of the same packet within a frame and each transmission carries a pointer to the slots where the other repetitions took place. On the receiving side, initially all transmission that occurred in a singleton slots are resolved.
Subsequently, successive interference cancellation is executed based on the information carried in the pointers, as illustrated in Fig.~\ref{fig:SIC}. By this approach the throughput of the scheme increases; for the simple scenario when each user performs two repetitions in randomly selected slots of the frame \cite{CGH2007}, the throughput increases to $T \approx 0.55$ for \emph {logical} load\footnote{We note that in SA with repetitions one should distinguish between the logical and physical load, where the former includes only ``unique'' transmissions and the latter includes includes all repetitions.} $G\approx0.6$.

In \cite{L2011} it was noted that the execution of SIC within the framed SA framework resembles the execution of the iterative belief-propagation (BP) decoding
on erasure channel, enabling the application of theory and tools of codes-on-graphs.
Following this insight, the author analyzed the convergence of the SIC using and-or tree arguments \cite{LMS1998} and obtained optimal repetition strategies in terms of maximizing throughput of the scheme.
It was shown that the optimal repetition strategies follow the same guidelines used for encoding of left-irregular LDPC codes.
In the asymptotic case when the number of users tends to infinity, both (logical) $G$ and $T$ tend to 1.
Nevertheless, for the optimal performance, the number of slots in the frame is determined by the number of transmissions.

Another important class of codes with advantageous erasure-correcting properties are rateless codes \cite{L2002,S2006}.
Rateless codes do not have a priori fixed code rate - the transmitter produces encoded symbols until a feedback signals that the message is decoded.
Using the ideas analogous to rateless coding, a \emph{frameless} SA scheme was introduced in \cite{SPV2012}; we briefly review it in the following subsection.

\subsection{Frameless ALOHA}
\label{sec:FRALOHA}

The length of the contention round in frameless ALOHA is not a priori fixed and the round lasts until the BS decides to end it by transmitting a notification to the contending users.
The users are synchronized both on the round and the slot basis, implying that the numbering sequence of the slots within the round is a common knowledge across users. %The synchronization can be achieved by a beacon that is sent by the BS prior to the start of the contention round.
In each slot of the contention round users attempt to access with a predefined probability, termed slot access probability, which is the same for all users and for a given slot. The slot access probability $p_m$ for slot $s_m$ is
\cite{SPV2012}:
\begin{equation}
p_m = \frac{\beta_m}{N} 
\end{equation}
where $\beta_m$ is the expected number of interfering transmissions in slot $s_m$, denoted also as the expected slot \emph{degree}, and $N$ is the number of users.
In the general case, $\beta_m$ is a function of the slot number $m$.
As shown in \cite{SPV2012}, the above access method results in a LT-like distribution \cite{L2002} of user transmissions over the slots of the round.
This distribution can be optimized using standard tools from the codes-on-graphs theory.
Coupled with SIC, for the realistic number of users the proposed access method enables throughputs higher the ones reported in \cite{L2011}.

As shown in \cite{L2002} and elaborated in \cite{SPV2012}, in order to resolve all user transmissions with a probability that tends to 1 when the number of slots tends to $N$, the average slot degree should scale as $O(\log N)$.
This requirement cannot be met in practice even for moderately high values of $N$, as it would adversely affect the performance of the interference cancellation algorithm.
Therefore, a contention round in frameless ALOHA ends when a sufficiently high fraction of user transmissions has been resolved, which can be achieved with a relatively low average slot degrees.
The feedback that signals the end of the round is sent by the BS, however, unlike in the framed ALOHA case, the time instant at which the feedback arrives is a priori not known and it adapts to the contention process.
The users with unresolved transmissions continue to contend in the following rounds, in the same way as in other SA schemes.

\subsection{Estimating the number of users}

The optimal operation of any variant of SA requires the knowledge of the number of users $N$.
In many practical situations, $N$ is a priori unknown and/or it varies over time, thus its estimate has to be obtained.
The estimation is regularly based on the observation of how many idle, singleton and/or collision slots have occurred within a frame.

In an early work \cite{Sch1983} set in the framework of framed SA, the estimate of the current number of backlogged users is obtained using the number of collision slots in the current frame.
The estimate determines the length of the next frame, with an aim of throughput maximization.
An approach suitable for the cases in which the frame length is fixed is suggested in \cite{RALRCP2005}, where the estimate of the number of contending users is based on the number of singleton slots within the frame.
The estimate determines the probability that users choose to contend in/skip the next frame such that the expected number of contending users is tuned to the frame length and thus the expected throughput is maximized.

The estimation of the number of users is in particularly of interest in RFID systems \cite{V2002,KN2006,EL2010}.
In \cite{V2002} a simple estimator is derived that combines the number of singleton and collision slots.
In \cite{KN2006} the estimation based either on the number of idle or collision slots, depending on the frame length and the number of RFID tags.
The algorithm presented in \cite{EL2010} improves the approach of \cite{Sch1983}, enhancing the estimation by using the number of the observed singleton slots as well.

All the mentioned estimators assume that the round length is set before it commences and that the users attempt (at most) a single transmission within it; this approach is unfitting for the frameless ALOHA scenario.
Instead, we use an approach in which the round length is not a priori set and users probabilistically attempt to transmit on a slot basis with predefined slot access probabilities, as elaborated in Section \ref{sec:estimation}.

\section{System model}
\label{sec:model}

\begin{figure}[tbp]
	\begin{center}
\includegraphics[width=0.92\columnwidth]{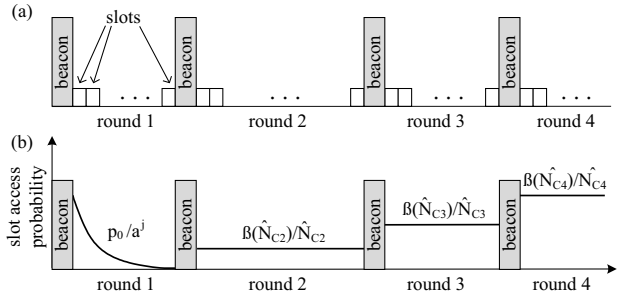}
	\end{center}
\caption{(a) Rounds of proposed protocol. (b) Variation of slot access probability within and over rounds.}
	\label{fig:phases}
\end{figure}

We consider the following setup. The network consists of $N$ users, where $N$ is lower- and upper-bounded but a priori unknown, which contend to access the BS with their uplink transmissions.
We focus on the resolution of a batch user arrival, however, the presented scheme can be easily be extended to the scenarios with continuous user arrivals; we outline the possibilities for this in Section~\ref{sec:conclusion}.

The access protocol operates in rounds that are divided into equal-duration slots, as depicted in Fig.~\ref{fig:phases}(a).
The beginning and the end of each round is indicated by a downlink beacon, sent by the BS.
The beacons are also used to synchronize the users both on the round and slot basis, to announce the function that determines the behavior of the slot access probability in the next round, as well as to acknowledge previously resolved users.
All rounds are both used for the estimation of the number of users $N$ and resolution of their transmissions; however, the emphasis in the initial round is on obtaining a sufficiently reliable estimation of $N$.
In the subsequent rounds the emphasis is on the contention resolution, while the estimation is being progressively improved.

Let $\hat{N}$ denote the current estimate of $N$.
Let $s_{ij}$ we denote $j$-th slot of $i$-th round, $|s_{ij}|$ the corresponding slot degree (i.e., the number of colliding user transmissions in slot $s_{ij}$) and $p_{ij}$ the slot access probability for $s_{ij}$.
The total number of resolved users after the $i$-th round (which becomes known as the protocol operation progresses) is denoted by $N_{Ri}$; the estimate of the number of contending users in the $i$-th round is thus $\hat{N}_{Ci}=\hat{N} - N_{R,i-1}$.
By default, we assume $N_{R0}=0$.
%In further text, we elaborate the protocol operation in details. 

\subsection{Estimation algorithm}
\label{sec:estimation}

The BS observes each slot $s_{ij}$ as being either an idle, a singleton or a collision slot.
The probability mass function of this observation, conditioned on the number of users $N$, is given by:
\begin{align}
f (s_{ij} | n ) =  \left\{
    \begin{array}{ll} 
      ( 1-p_{ij} )^{n_{Ci}}  &  \text{if } |s_{ij}| = 0, \\
      n_{Ci} p_{ij} ( 1-p_{ij})^{n_{Ci} - 1} & \text{if } |s_{ij}| = 1, \\
      1 - ( 1-p_{ij} )^{n_{Ci}} - \\ - n_{Ci} p_{ij} ( 1 - p_{ij} )^{n_{Ci} - 1} & \text{if } |s_{ij}| > 1.  \\
    \end{array}
  \right.
\end{align}
where $p_{ij}$ is the slot access probability for $s_{ij}$ and $n_{Ci} = n - n_{R,i-1}$ is the number of contending users.\footnote{For the sake of clarity, we note that variables $n$, $n_{Ci}$, $n_{R,i-1}$ correspond to the random variables $N$, $N_{Ci}$ and $N_{R,i-1}$, respectively.}

We perform maximum likelihood estimation (MLE) of $N$ using sequence of observations $\{s_{ij}\}$.
Due to the independence of slots, the MLE of $N$ is given by:
\begin{align}
	\label{eq:prod}
  \hat N & = \arg \max_{n} \prod_{i,j} f(s_{ij} | n ) = \arg \max_{n} \sum_{i,j} \ln f( s_{ij} | n ),
\end{align}
and it is obtained by solving the following equation for $n$:
\begin{align}
 \frac{\partial }{\partial n} \left( \sum_{i,j} \ln f (s_{ij} | n) \right) = \sum_{i,j} \left[ \frac{\partial }{\partial n}\ln f (s_{ij} | n ) \right]= 0,
\label{eq:maximize}
\end{align}
where $\frac{\partial} {\partial n} \ln f( s_{ij} | n )$ is:
\begin{align}
  & \frac{\partial} {\partial n} \ln f( s_{ij} | n ) = \nonumber \\
  \label{eq:derlog}
   & = \left\{
    \begin{array}{ll} 
      \ln ( 1- p_{ij} )  &  \text{if } |s_{ij}| = 0, \\
      \frac{1}{n_{Ci}} +  \ln ( 1-p_{ij} ) & \text{if } |s_{ij}| = 1, \\
      \frac{( 1- p_{ij} )^{n_{Ci}}[ 1 + \ln ( 1 - p_{ij} ) ( \frac{1}{p_{ij}} + n_{Ci} - 1 ) ]}
      { 1 - \frac{1}{p_{ij}} + (1 - p_{ij} )^ {n_{Ci}}  ( \frac{1}{p_{ij}} + n_{Ci} - 1 ) } & \text{if } |s_{ij}| > 1.  \\
    \end{array}
  \right.
\end{align}
Combining \eqref{eq:maximize} and \eqref{eq:derlog} yields:
\begin{align}
\label{eq:full}
& \sum_{s_{ij}\in{\mathcal{O}_0} \cup \mathcal{O}_1} \ln ( 1- p_{ij} ) + \sum_{s_{ij}\in \mathcal{O}_1} \frac{1}{n_{Ci}} + \nonumber \\
& \sum_{s_{ij}\in \mathcal{O}_C} \frac{( 1- p_{ij} )^{n_{Ci}}[ 1 + \ln ( 1 - p_{ij} ) ( \frac{1}{p_{ij}} + n_{Ci} - 1 ) ]}
      { 1 - \frac{1}{p_{ij}} + (1 - p_{ij} )^ {n_{Ci}}  ( \frac{1}{p_{ij}} + n_{Ci} - 1 ) } = 0,
\end{align}
where $\mathcal{O}_0$, $\mathcal{O}_1$ and $\mathcal{O}_C$ are sets of observed idle, singleton and collision slots, respectively.
Eq. \eqref{eq:full} is solved for $n$ using the fact that $n_{Ci} = n - n_{R,i-1}$\footnote{The solution domain for $n$ is a priori set to $n > n_{R,i-1}$.}; this can be done efficiently using fast numerical root-finding methods like Brent's method \cite{B1973}. 

As evaluation of \eqref{eq:full} requires an iteration over all observed slots, and as the numerical root-finding methods also implies a separate iterative procedure, the overall evaluation complexity can be high.
However, it is reasonable to assume that the BS has enough processing power to address this challenge.

\subsection{Protocol operation in the initial round}

The slot access probability in the first round is given as:
\begin{align}
\label{eq:1st}
  p_{1j} = \frac{p_0}{\alpha^{j}},
\end{align}
where $p_0 \leq 1$ and $\alpha>1$ are suitably chosen constants.
The motivation behind (\ref{eq:1st}) is to progressively decrease slot access probability until very low values are reached, see Fig.~\ref{fig:phases}(b), and in such way cover the interval in which the slot states will transient from collision to idle.
The information on the number of contending users $N$ is ``contained'' in this transient interval, as collision/idle slots alone can not provide information on which the estimate will be made \cite{KN2006}.

The initial slot access probability $p_{11}=p_0$ should be chosen such that long (and uninformative) sequences of collision slots are avoided in the beginning of the initial round.
On the other hand, the choice $\alpha$ determines how fast the transient interval is traversed, reaching the point where only idle slots
appears, and hence $\hat{N}$ can not be further refined (implying that the round should be terminated).
The parameter $\alpha$ should be chosen such that a compromise between a fast and a reliable estimation is achieved - large $\alpha$ implies a short round length, but also a short length of the transient interval, which adversely affects the reliability of the
estimation.
We note that the actual choice of $p_0$ and $\alpha$ depends on the expected range of the number of users $N$; these parameters are set before the initial round starts, and, since there is no feedback until the end of the round, they can not be adjusted on the way.

The initial round should be terminated when only idle slots start to appear, bringing no improvement to the estimation.
A simple rule is to end the round after observing $K$ consecutive idle slots, where $K$ is suitably chosen constant.

The estimate $\hat{N}$ is refined after the every slot of the initial round, i.e., $\hat{N}$ is updated using \eqref{eq:full} after each newly observed slot.
Also, during the course of the initial round, some of the users (i.e., user transmissions) will already be resolved using SIC, as described in
Section~\ref{sec:background}.
The output of the initial round is $\hat{N}_{C2} = \hat{N} - N_{R1}$, where $N_{R1}$ is the number of resolved users in the initial round.

\subsection{Protocol operation in the subsequent rounds}
\label{sec:contention}

After the initial estimate on the number of contending users is obtained, the main purpose of the subsequent rounds is to resolve user transmissions.
As shown in \cite{SPV2012}, a simple and efficient approach is to set the slot access probability to be constant within the round:
\begin{align}
\label{eq:2nd}
  p_{ij} = p_i = \frac{ \beta (\hat{N}_{Ci} )}{\hat N_{Ci}},
\end{align}
for every round $i \geq 2$, where $\beta (\hat{N}_{Ci} )$ is chosen such that throughput of the access scheme is maximized for $\hat{N}_{Ci}$ \cite{SPV2012}.

As outlined in Section~\ref{sec:FRALOHA}, a round is terminated when a predefined fraction $\gamma(\hat{N}_{Ci})$ of the estimated number of contending users $\hat{N}_{Ci}$ is resolved.
In \cite{SPV2012}, the fraction $\gamma(\hat{N}_{Ci})$ is chosen such that the expected maximum throughput is reached.
However, here we have to include the effect of the estimation errors when deciding on $\gamma(\hat{N}_{Ci})$; particularly as the choice of too high $\gamma(\hat{N}_{Ci})$ coupled with an overestimation could lead to deadlock situations.
We outline the principles for selecting appropriate $\gamma(\hat{N}_{Ci})$ in more details in Section~\ref{sec:subseq_res}.

An exception to the above stopping criterion has to be made in the final round, when all the remaining contending users should be eventually resolved.
A round should be designated as the final one when the number of contending users becomes sufficiently low; more details on this are given in Section~\ref{sec:subseq_res}.

\section{Results}
\label{sec:results}

In this section we present simulation results for the proposed protocol.
We assume that the number of users $N$ is within range $[100,10000]$ and we tune the parameters of the protocol accordingly.
We have performed only MAC-layer simulations, abstracting the physical layer issues (propagation, modulation, coding, synchronization, possible capture effects...).
The SIC algorithm was implemented using standard iterative BP decoder \cite{L2002}.
However, as outlined in \cite{L2011}, despite the simplifications of the approach, the obtained results closely match the ones obtained by the more comprehensive and complex simulations, at least for moderate to low packet-loss rates.
All results are obtained using 5000 simulation repeats for each set of the parameters' values.

\subsection{Choosing the parameters of the initial round}

\begin{figure}[tbp]
	\begin{center}
\includegraphics[width=0.9\columnwidth]{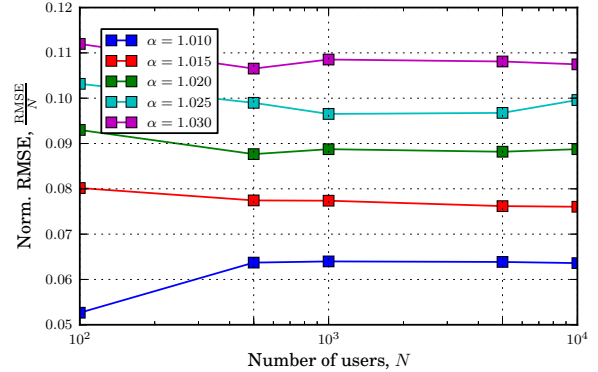}
	\end{center}
\caption{Normalized RMSE of the estimate after the initial round for different values of $\alpha$, when $p_0=0.047$ and $K=6$.}
	\label{fig:std}
\end{figure}

Parameter $p_0$ should be chosen such that the collision slots appear at the beginning of the initial round with a sufficiently high probability $P_C$, with respect to the minimal number of contending users $N_{\min}$.
For the targeted range of $N$, where $N_{\min}=100$, $p_0$ should be chosen such that:
\begin{equation}
1 - (1 - p_0)^ {N_{\min}} -  N_{\min} p_0 (1 - p_0)^ {N_{\min} - 1 } \geq P_C. 
\end{equation}
Simulations showed that setting $P_C \geq 0.95$ does not affect the accuracy of the estimation; for $P_C=0.95$ and $N_{\min}=100$ we get $p_0\approx0.047$.
Regarding the choice of the number of observed consecutive idle slots $K$ after which the initial round should be terminated, the simulations showed that for $K \geq 6$ the estimate $\hat{N}$ can not be refined any more, as measured by its root-mean square error (RMSE).
In other words, the choice $K=6$ seems to be a near-to-optimal one.

\begin{figure}[tbp]
	\begin{center}
\includegraphics[width=0.9\columnwidth]{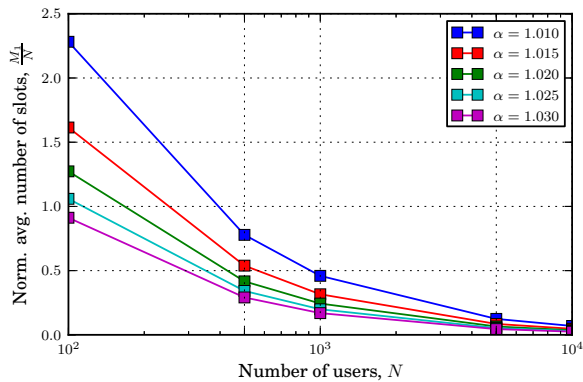}
	\end{center}
\caption{The length of the initial round (in slots), normalized by the number of users $N$, for different values of $\alpha$ and when $p_0=0.047$ and $K=6$.}
	\label{fig:length}
\end{figure}

Further, the simulations showed that the proposed estimator is not biased and that the probability density function (pdf) of $\hat{N}$ closely matches a Gaussian pdf that is centered at $N$ and has standard deviation equal to the simulated RMSE\footnote{We leave the analytical proof of this result for future work.}.
Fig.~\ref{fig:std} shows normalized RMSE of $\hat{N}$, as function of $N$ and for different values of $\alpha$.
As it can be observed, normalized RMSE does not change significantly with $N$.
Fig.~\ref{fig:length} shows the average number of slots of the initial round $M_1$ normalized with $N$, for the same values of $N$ and $\alpha$.
For low $N$ the ratio $M_1/N$ is rather high, which adversely affects the total throughput\footnote{The same holds for all the related estimators - to estimate low number of users with high confidence requires disproportionally high number of slots.} (see Section~\ref{sec:perf}).    
Comparing Figs.~\ref{fig:std} and \ref{fig:length}, the trade-off between the accuracy of the estimation and the initial round length becomes clear.
In further text we demonstrate the performance of the proposed protocol for $\alpha = 1.02$.

\subsection{Choosing the parameters of the subsequent rounds}
\label{sec:subseq_res}

The parameters of the subsequent rounds $\beta({\hat{N}_{Ci}})$ and $\gamma({\hat{N}_{Ci}})$ are selected using simple rules outlined below. 
We set $\beta({\hat{N}_{Ci}})=2.9$, as this value represents a simple but effective approximation that is based on the results given in \cite{SPV2012}.
Besides providing for close-to-optimal performance of the collision resolution through the SIC, this choice of $\beta({\hat{N}_{Ci}})$ has advantageous properties for the proposed estimation algorithm as it provides a useful ``mixture'' of idle, singleton and collision slots in the subsequent rounds (see Section~\ref{sec:estimation}).

We set $\gamma(\hat{N}_{Ci})=0.8$ based on the following reasoning.
For $\alpha=1.02$, after the initial round RMSE~$\approx 0.09N$ (see Fig.~\ref{fig:std}) and thus $\hat{N} \leq 1.25N$ and $\hat{N}_{Ci} \leq 1.25 N_{Ci}$ with confidence greater than $0.99$.
This implies that $\gamma(\hat{N}_{Ci}) \hat{N}_{Ci} = 0.8 \hat{N}_{Ci} < N_{Ci}$ in more than 99\% of the cases, ensuring that the round is not indefinitely protracted by waiting for the non-existing (i.e., overestimated) users to become resolved.
To address the remaining unlikely events of overestimation, we introduce an auxiliary stopping criterion - the round is forced to end after $2\hat N_{Ci}$ slots have elapsed.

We note that setting $\gamma(\hat{N}_{Ci})=0.8$ does not necessarily imply that only up to $0.8N_{Ci}$ users are resolved within $i$-th round.
Namely, it is well-known that the iterative BP decoder (i.e., SIC in our framework) for properly designed degree distribution exhibits a threshold-like behavior \cite{M2005}; in the beginning the number of resolved users increases slowly as the round progresses, however, at a certain point there is a surge in the number of resolved users and the fraction of resolved users becomes rather close to 1.
In other words, in a matter of a single slot the resolved fraction of $N_{Ci}$ substantially surpasses $0.8$,\footnote{Which also denotes that this slot is going to be the last slot of the round.} provided that the actual slot access probability $\beta(\hat{N}_{Ci})/\hat N_{Ci}$ is sufficiently close to the desired $\beta(N_{Ci})/N_{Ci}$.
This is equivalent to the condition that $\hat{N}_{Ci}$ is sufficiently close to $N_{Ci}$; the performed simulations for the chosen $p_0$ and $\alpha$ showed that this condition is satisfied with high probability already after the initial round and it progressively improves in the subsequent rounds, as displayed in Section~\ref{sec:perf}.
 
A round $i$ is designated as the final one when after its completion both of the following two conditions are met: (1) $p_i > 0.5$ and (2) $|\hat{N} - N_{ri}| < 1$.
The first condition ensures that there are sufficiently many rounds before the final one and thus the estimate becomes very reliable (see Section~\ref{sec:perf}).\footnote{The simulations showed that the probability of not resolving all users when using the proposed criteria is minuscule. Nevertheless, an extra round/rounds can be appended with an aim of verifying that all users have been resolved.}

\subsection{Performance of the proposed protocol}
\label{sec:perf}

\begin{figure}[tbp]
	\begin{center}
\includegraphics[width=0.9\columnwidth]{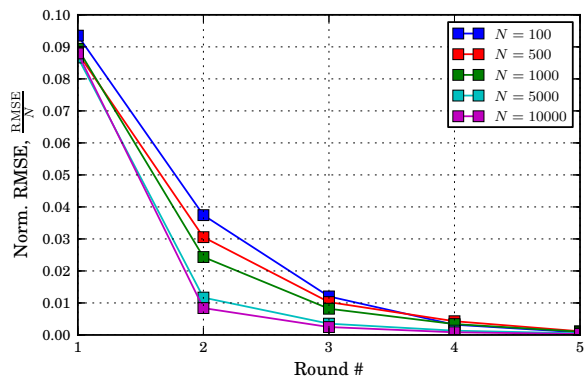}
	\end{center}
\caption{The normalized RMSE as function of the number of the rounds, when $\alpha=1.02$, $p_0=0.047$ and $K=6$; the estimate is updated only after each round and no backtrack is used.}
	\label{fig:est}
\end{figure}

With respect to updating of the initial estimate, we assess the performance of the following protocol variants:
\begin{itemize}
	\item $\hat{N}$ is updated after every round $i$, $i \geq 2$.
	\item $\hat{N}$ is updated both when number of the slots in the every round $i$ reaches $0.5\hat{N}_{Ci}$ and after every round $i$, $i \geq 2$.
	\item $\hat{N}$ is updated after every slot.  
\end{itemize}
Also, with respect to the execution of the SIC algorithm, we examined the following two cases:
\begin{itemize}
	\item The SIC is executed only on the slots that occur in the current round (further denoted as ``no backtrack'').
	\item The SIC is executed on all previously occurred slots (denoted as ``backtrack'').
\end{itemize}
We note that in both of the variants, we executed the SIC only on slots $s$ for which $|s| \leq 10$ (i.e., only the slots that contain ten or less interfering transmissions are considered exploitable for the SIC), as it is reasonable to assume that the interference cancellation can not successfully operate on high-degree slots.
In this way the usage of high-degree slots at the beginning of the initial round for the SIC is avoided and the obtained results can be considered as more realistic.

We are interested both in the overall estimation and collision-resolution performances.
The estimation performance is expressed through root-mean-square error (RMSE) of $\hat{N}$.
The collision-resolution performance is measured using average throughput of the scheme $T$ after all users have been resolved; $T=N/M$, where $M$ is the average of the total number of slots needed to resolve $N$ users. 

Fig.~\ref{fig:est} shows the normalized RMSE as function of the number of the rounds, for the case when $\hat{N}$ is updated only after each round and no backtrack.
This could be considered as the worst case performance, as the estimate gets updated least frequently. 
It can be observed from that as the number of round increases, RMSE drops to rather small values, implying that $\hat{N}$ becomes very accurate.
The simulations also demonstrated that all the other variants of the estimation updating/SIC execution have only modestly better estimation performance.

\begin{figure}[tbp]
	\begin{center}
\includegraphics[width=0.9\columnwidth]{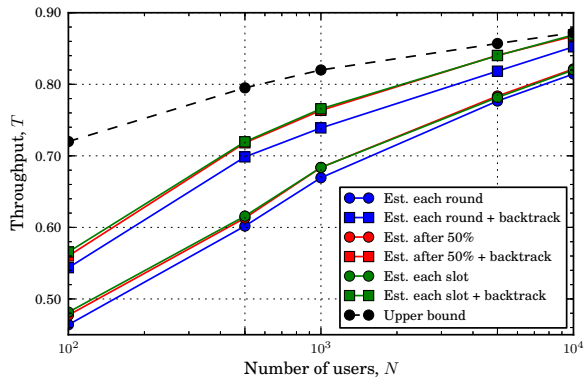}
	\end{center}
\caption{Throughput of the proposed protocol when $\alpha=1.02$, $p_0=0.047$ and $K=6$.}
	\label{fig:throughput}
\end{figure}

Fig.~\ref{fig:throughput} shows the throughput for all the combinations of estimation updating and SIC execution.
Obviously, using backtrack leads to notably higher throughputs.
Further, the variants in which $\hat{N}$ is updated more frequently show improved performance; however, there is no difference between the case when the estimate is updated after each slot and the case when it is updated just once during the round, when number of elapsed slots reach half of the estimated number of currently contending users.
As the former approach introduces significantly higher computation burden, it is fair to conclude that the latter approach is a better choice.
Comparing the above results with the upper bound derived using the results presented in \cite{SPV2012}, it is clear that as $N$ increases, the obtained throughputs tend to their limits.

\section{Discussion and conclusions}
\label{sec:conclusion}

A comprehensive approach to optimization of the parameters of the proposed scheme, i.e., $p_0$, $\alpha$, $K$, $\beta(N_C)$ and $\gamma(N_C)$, should take all of them into account jointly with an aim of the total throughput maximization.
However, the corresponding analysis would inevitably be intractable, due to the multidimensionality of the problem and mutual interdependencies of the parameters.
The simplified approach used in the paper demonstrated the potential of the proposed scheme and obtained results comparable to the upper bounds given in \cite{SPV2012}; a more complex optimization could be worthwhile for lower number of users.
Nevertheless, one should be aware that the throughput losses for low $N$ are mainly due to the length of the round where the initial estimate has to be obtained, which would be the case for any estimation algorithm of the type.

When comparing the results for the variants with backtrack/no backtrack, the former show improved performance.
On the other hand, the memory requirements in both cases are comparable, as the length of the second round (when most of the users become resolved) constitutes the highest portion of the overall length.
 
The integral part of the scheme are the beacons sent by the BS with a purpose of ending the current round/starting the next round with the updated slot access probability.
In FDD systems, the beacons can be broadcasted using the downlink channel.
In TDD systems, the beacons could be given precedence over the user transmissions using the strategies similar to the ones based on SIFS/DIFS in IEEE 802.11 and related wireless standards, thus avoiding collisions.

Finally, the proposed protocol could easily be extended for the scenarios with continuous arrivals, in the same way that related framed SA protocols operate \cite{R1975,Sch1983,RALRCP2005}.
The protocol operation in this case would be divided into phases; all new arrivals occurring during a phase are backlogged, waiting for the next phase to start.
In this case, provided that there is a correlation between the number arrivals over the phases\footnote{E.g., the constant mean arrival rate for the Poisson arrivals.}, the proposed estimating procedure could be straightforwardly modified to exploit it.
We leave this for further work.

\section*{Acknowledgment}

The research presented in this paper was supported by the Danish Council for Independent Research (Det Frie Forskningsr\aa d) within the Sapere Aude Research Leader program, Grant No. 11-105159 ``Dependable Wireless Bits for Machine-to-Machine (M2M) Communications''.


\end{document}